\documentclass[sigconf]{acmart}

\usepackage[english]{babel}
\usepackage{blindtext}

\renewcommand\footnotetextcopyrightpermission[1]{} 
\setcopyright{none}

\acmDOI{}

\acmISBN{}

\acmConference[]{}
\acmYear{}
\copyrightyear{}

\acmPrice{}

\usepackage{xspace}
\usepackage{url}
\usepackage{multirow}
\usepackage{amsmath}
\usepackage[normalem]{ulem}
\usepackage{xcolor}
\usepackage{wrapfig}
\usepackage{graphicx}
\usepackage{caption}
\usepackage{float}

\usepackage{cancel}
\renewcommand{\sout}[1]{}
\newcommand{\newadd}[1]{\textbf{\color{teal} #1}}
\renewcommand{\newadd}[1]{#1}

\DeclareMathOperator*{\argminA}{arg\,min}
\newcommand{\theo}[1]{\textbf{\color{red} Theo: #1}}
\renewcommand{\theo}[1]{}

\newcommand{\usama}[1]{\textbf{\color{blue} Usama: #1}}
\renewcommand{\usama}[1]{}
\newcommand{\reword}[1]{\textbf{\color{purple} #1}}
\renewcommand{\reword}[1]{#1}

\newcommand{\note}[1]{}

\newcommand{\BO}{Bayesian optimization\xspace}

\newcommand{\AB}{Apache Benchmark\xspace}
\newcommand{\sysname}{ConfigTron\xspace}

\newcommand{\obj}[1]{\textbf{\color{RubineRed} Points to focus: #1}}
\renewcommand{\obj}[1]{}

\newcommand{\api}{ConfigTron-API\xspace}
\newcommand{\agent}{Configuration Agent\xspace}
\newcommand{\agents}{Configuration Agents\xspace}
\newcommand{\manager}{Configuration Manager\xspace}

\newcommand{\parab}[1]{\vspace*{0.05in}\noindent\textbf{#1}}
\newcommand{\configuration}{configuration\xspace}

\newcommand{\managers}{managers\xspace}

\usepackage{graphicx,subcaption}
\usepackage{enumitem}

\begin{document}
	\date{}
	\title{\sysname: Tackling network diversity with heterogeneous configurations}
	\author{Usama Naseer\qquad Theophilus Benson\\Brown University}

	\begin{abstract}

The web serving protocol stack is constantly changing and evolving to tackle technological shifts in networking infrastructure and website complexity.  
As a result of this evolution, the web serving stack includes a plethora of protocols and configuration parameters that enable the web serving stack to address a variety of realistic network conditions.  Yet, today, most content providers have adopted a ``one-size-fits-all'' approach to configuring the networking stack of their user facing web servers (or at best employ moderate tuning), despite the significant diversity in end-user networks and devices.

In this paper, we revisit this problem and ask a more fundamental question: Are there benefits to tuning the network stack? If so, what system design choices and algorithmic ensembles are required to enable modern content provider to dynamically and flexibly tune their protocol stacks. 
We demonstrate through substantial empirical evidence that this ``one-size-fits-all'' approach results in sub-optimal performance and argue for a novel framework that extends existing CDN architectures to provide programmatic control over the configuration options of the CDN serving stack. We designed \sysname a data-driven framework that leverages data from all connections to identify their network characteristics and learn the optimal configuration parameters to improve end-user performance.
\sysname uses contextual multi-arm bandit-based learning algorithm to find optimal configurations in minimal time, enabling a content providers to systematically explore heterogeneous configurations while improving end-user page load time by as much as 19\% (upto 750ms) on median. 

\usama{Reminder: revisit abstract and conclusion.}

\end{abstract}
	\maketitle
	
	
	\vspace{-0.1in}
\section{Introduction}

Web page performance significantly impacts the revenue of content service providers (CSP) (e.g., Facebook or Google) and content distribution networks (CDNs), with recent studies showing that a 400ms increase in page load times (PLT) can reduce revenue by 4.3\%~\cite{bing,brutlag2009speed}.
Yet, uniformly improving web performance is becoming increasingly challenging because of the growing disparity in the network conditions (bandwidth, RTT and loss rates) \cite{singh2015flexiweb, wang2014speedy, al2011overclocking, erman2015towards} and devices used by end-users~\cite{ruamviboonsuk2017vroom, nsdi:wprof, www:wprofx}.
To address this disparity and improve quality of experience (QoE), the networking community is constantly developing new protocols and configuration suggestions for user-facing web servers (AKA, edge servers), e.g., PCC~\cite{dong2015pcc}, Vivace~\cite{vivace}, and BBR~\cite{card2017wellbbr} at the congestion layer (L4) and  QUIC~\cite{quic}, SPDY~\cite{spdy} and HTTP2 at the application layer (L7). 


The optimal choice of protocol and parameters is contingent on the network infrastructure~\cite{wang2014speedy,erman2015towards,karlsson2007tcp, singh2015flexiweb, al2011overclocking, tcpwise, atc:pantheon, ruffy2018iroko, jay2018internet, li2018qtcp}, website complexity~\cite{butkiewicz2011understanding, nsdi:klotski, netravali2016polaris, wang2016speeding, nsdi:wprof}, and end-user device~\cite{ahmad2016view, ruamviboonsuk2017vroom, www:wprofx}.  Furthermore, innovations along any one of these three dimensions will lead to the invention of new protocols and the re-evaluation of default configuration parameters.  Although different regions and ISPs leverage radically different networking infrastructure and mobile devices \cite{ahmad2016view}, a majority of CSPs continue to employ a ``one-size-fits-all'' approach to configuration, which results in sub-optimal performance and high tail-latency in certain regions~\cite{al2011overclocking, tcpwise, wang2014speedy, erman2015towards}. 
	

	 \begin{figure}[t!]
		\begin{center}
			\includegraphics[width=2.4in]{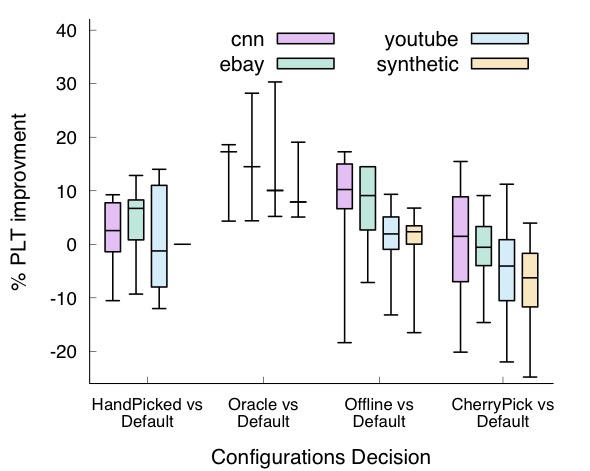} 
			\caption{PLT comparison}
			\label{f:bestvsdefault}
		\end{center}
	\end{figure}

\subsection{Tuning  Configurations Today}
Most attempts to tackle this growing diversity involve manual analysis of the performance of different configuration options in specific regions or on specific websites. Unsurprisingly, while several CSPs expose configuration knobs to their costumers~\cite{vcl-fastly}, many customers do not take advantage of them due to manual labour required~\cite{vcl-fastly} to tune these knobs. Alternatively, when operators tune their stack the focus on a limited set of knobs~\cite{akamai-knobes}. 

In this paper, our focus is on tuning a board set of configuration knobs across the transport (e.g., congestion control algorithm) and  application layers (e.g., HTTP version). Table~\ref{t:configurations} highlights the knobs currently being tuned by our system. In addressing this problem, we focus on first understanding the challenges associated with tuning and quantify benefits of tuning before we explore the systems and algorithmic considerations introduced by fined-grained cross layer tuning.

\begin{table}[tb!]
	\begin{center}
		\begin{scriptsize}
			\begin{tabular}{|p{27pt}|p{55pt}|p{43pt}|c|}
				\hline 
				\textbf{Layer}&\textbf{Protocol Options} & \textbf{Default Value} & \textbf{Values tested} \\ \hline
				\multirow{5}{*}{TCP}&congestion\_control & Cubic & Cubic, Reno, Vegas, BBR \\ \cline{2-4}
				&initcwnd & 10 MSS & {1, 4, 10, 16, 20, 30} \\ \cline{2-4}
				&slow\_start\_after\_idle & 1 & 0, 1 \\ \cline{2-4}
				&low\_latency & 0 & 0, 1 \\ \cline{2-4}
				&autocorking & 1 & 0, 1 \\ \cline{2-4}
				&pacing & pfifo\_fast & pfifo\_fast, fq  \\ \hline
				{Web App} & HTTP Protocol & 1.1 & 1.1, 2\\ \hline
			\end{tabular}
		\end{scriptsize}
	\end{center}
	\caption{Web stack configuration parameters.}
	\vspace{-0.45in}
	\label{t:configurations}
\end{table}

To illustrate the difficulty of tuning server configurations, in Figure~\ref{f:bestvsdefault}, we compare page load times of CDN servers when configured using popular tuning techniques. Specifically, \textit{Bayesian Optimization}~\cite{pelikan1999boa} (BO) (e.g., CherryPick~\cite{alipourfard2017cherrypick}) a popular statistical technique used for tuning systems configurations~\cite{duan2009tuning,herodotou2011starfish, van2017automatic,alipourfard2017cherrypick,fb-bo}, operator \textit{hand-tuned} configurations (discussed in~\S~\ref{s:motivation}), closed-loop \textit{offline-learning}, and an \textit{oracle} technique. For default configurations, we used Linux's default (Table~\ref{t:configurations}). 

Hand-tuned configurations are manually selected and are thus, naturally, coarse-grained: unsurprisingly, we observe that while hand-tuned improves median it fails to provide optimal performance across varying network conditions hence the negative improvement values.

\BO, aims to quickly discover ``good'' configuration; while fine-grained, this approach is relatively static, it does not re-evaluate old choices and is thus unable to adapt to network dynamics. In Figure~\ref{f:bestvsdefault}, we observe effects of this rigid behavior with wildly fluctuating performance.

Lastly, we explore an offline model which learns on traces from prior days, and applies the learned model on network connections for the next day. Offline model is fine-grained but with limited dynamicity: the trained model is unable to react to realtime issues. Unfortunately, due to the high-dimensionality of the Internet's dynamics these realtime issues are the norm not the exception~\cite{jiang2016cfa, sigcomm:metis}. We observe in Figure~\ref{f:bestvsdefault}, that offline performs closest to the optimal (oracle's behavior) but still falls short because of its inability to react in real time.

Our brief analysis of modern approaches to tuning highlights the need for a highly dynamic and fine-grained approach to tuning the configurations of the web server's network stack.  An approach which is able to adapt, in real-time, to changes within the underlying network.

\begin{table*}[!htbp]
	\begin{center}
		\begin{scriptsize}
			\begin{tabular}{|c|c|c|c|c|}
				\hline 
				\textbf{Layer } & \textbf{Option} & \textbf{Top configs. in NA (cross-CSP)} & \textbf{\% heterogeneity outside NA } & \textbf{Instance of heterogeneity observed} \\ \hline
				
				\multirow{5}{*}{HTTP (L7)} & HTTP version & H1.1(44.3\%), H2(55.7\%) & 4.7\% & NA H2 -$>$ Asia H1.1 \\ \cline{2-5}
				& Max header list size  & 16384 (100\%) & 0\% & None \\ \cline{2-5}
				& Header table size & 4096 (100\%) & 0\% & None \\ \cline{2-5}
				& Max concurrent streams & 100 (44\%), 128 (56\%) & 1\% &  NA 100 -$>$ EU 128 \\ \cline{2-5}
				& Initial window size & 65536 (71\%), 65535 (15\%), $>$1M (14\%) & 1.9\% & NA 1048576B -$>$ Asia 65535B \\ \cline{2-5}
				& Max frame size & 16,777,215 (81\%), 16384 (19\%) & 0\% & None \\ \hline
				
				\multirow{3}{*}{TCP (L4)} & ICW & \{10 (62\%), 4(20.5\%), 24(5.3\%)\} MSS & 6.9\% & NA 24 MSS -$>$ Asia 10 MSS \\ \cline{2-5}
				& RTO & \{0.3(9.2\%), 1(82.6\%), 3(8.2\%)\} sec & 2.3\% & NA 3s -$>$ EU 1s \\ \cline{2-5}
				& RWIN & \{29200(57.4\%), 14600(8.2\%), 42780(6.8\%)\} bytes & 3.6\% & NA 29200B -$>$ Asia 12960B \\ \hline
				
			\end{tabular}
		\end{scriptsize}
	\end{center}
	\caption{Heterogeneity in configs. across 5 regions}
	\label{t:ig}
	\vspace{-0.3in}
\end{table*}

\subsection{\sysname}

In this paper, we eschew the notion of a homogeneous approach to tuning web server configurations and instead argue for a ``curated'' approach to configuring the web server's network stack. In particular, we argue that edge servers should be configured and setup with the optimal protocols and \configuration parameters required to serve each of the incoming connections, e.g.,  \theo{can we replace this ICW examples with HTTP and congestion control} an edge server serving high loss, low bandwidth connections may employ a lower initial window size than one serving low loss, high bandwidth connections.

%
To this end, we argue for a simple but powerful server architecture that introduces flexibility into the network stack, enables reconfiguration and systematically controls configuration heterogeneity.
  %
We also introduce a contextual multi-armed bandit-based learning algorithm, an embodiment of domain specific insights, which tunes configuration in a principled manner to find optimal configurations in minimal time.
	%
Taken together our system design and learning algorithm enables a CSP/CDN to systematically explore heterogeneity in a dynamic and fine-grained manner while improving end-user performance in a principled manner. 

The design of \sysname faces several practical and interesting challenges:  
%
\begin{itemize} [topsep=0.1ex,leftmargin=2\labelsep,wide=0pt]
	\setlength\itemsep{0em}
	
	\item Non-stationary: network conditions are dynamic (changing every few minutes~\cite{zhang2001constancy, urvoy2005stationarity, lu2005characterizing, balakrishnan1997analyzing, jiang2016cfa}). Thus, \sysname must quickly and continuously learn optimal configurations.
	
	\item Non-Gaussian noise: Most CSPs focus on improving tail latency~\cite{tail-at-scale,dynamo,sigcomm:metis} which is often caused by non-Gaussian processes (e.g., last mile contention~\cite{sundaresan2016home}, mobile device limitations). Thus, \sysname must address non-Gaussian noise.
	 
		\item High-dimensionality: Content personalization, diverse devices and last mile connections~\cite{sundaresan2016home} introduce high dimensionality that make  offline closed loop approaches in-effective \cite{jiang2016cfa, nsdi:pytheas}\theo{citations from the CFA line of research?}. Thus, \sysname explore an online and scalable approach to tackle high dimensionality of internet conditions.
		
		\item High-cost data: Generating data for learning requires testing configuration and impacting user performance -- a very costly endeavor. Thus, \sysname must minimize impact to end users.
		

\end{itemize}

The key insight of \sysname is to simultaneously operate in two modes depending on the ``quality'' of the performance model: essentially, \sysname bootstraps model building intelligently and adaptively selects samples that speed up model convergence then, at steady state transitions to a greedy-mode that stochastically samples random points to iteratively improve performance:
\sysname tackles data-scarcity and high-data cost by clustering clients and sampling across the cluster.

\sysname uses a contextual multi-armed bandit designed explicitly to continuously learn an optimal configuration within a minimal number of exploration steps. 
Our ensemble fuses the stateful exploration of Gaussian-bandit (\S~\ref{s:bandits}) with the non-determinism of Epsilon-bandit: statefulness enables informed exploration of the configuration space while the randomness allows re-sampling of old configurations and continuous searching to identify optimal configuration. The re-evaulation provided by resampling enables \sysname to directly tackle non-Gaussian noise within the domain and the non-stationary property of the problem.  The data-collected by the ensembled is encoded in a decision tree -- which enables quick and easy classification but is also amenable to automatic generation of rules for the CDN server.

To demonstrate the benefits of \sysname, we conducted large-scale simulations to analyze the implications of various design choices on the performance and accuracy of \sysname. \reword{Our evaluations show that \sysname provides as much as 19\% (upto 750ms) reduction in PLT for the median case and a 36-80\% (900ms-1400ms) improvement in the tail latency. Given the recent arms race by CSPs to improve web performance, we believe that \sysname's modest performance improvements will result in significant revenue savings.}

\theo{cut the testbed out for now.}
\section{Empirical Study}
\label{s:motivation}

\obj{We did not observe region-specific configurations. Next, we want to show that why one-size fits all is a problem and what are its implications. To illustrate this point, we focus on few configurations examples such as ICW and CC. We also make the point TCP's like PCC or Remy are also limited in their ability to tackle different network conditions. We show this with the help of data from pantheon.}

In this section, we analyze CSPs to determine the extent of configuration tuning (\S~\ref{s:inspectorgadget}) that exists today and quantify the implications of configuration tuning (\S~\ref{s:config_study}).

\subsection{Probing web-servers configurations}
\label{s:inspectorgadget}

\usama{Assumption/claim: vantage point from AWS can exhibit the network-based heterogeneity. We don't mention any limitations here.}
We aim to understand if modern CSPs employ homogeneous configurations, as suggested by anecdotal evidence, or heterogeneous configurations to tackle diversity in Internet ecosystem. To this end, we developed a tool that combines and extends existing measurement tools~\cite{yang2014tcp, oliver-icw, icdcs:ig} to recognize modern protocol versions and inter-operate with TLS. Our tool actively probes web servers and infers configuration parameters by inspecting the packet-headers and the server's reaction to emulated network events (e.g., packet loss). We encourage the reader to peruse prior-work~\cite{yang2014tcp, oliver-icw, icdcs:ig} for further explanation.

We fingerprint configuration choices across three layers: application/L7 and transport/L4 for the Alexa top 1k websites. We compare the configurations for the same CSP across five different regions (North America, South America, Asia, Europe and Australia). Using North America (NA) as the baseline, Table~\ref{t:ig} presents the degree of heterogeneity (percentage of web servers that configure differently in NA vs other regions) observed for the Alexa top 1k. 


\textbf{Observation-1: Heterogenity across CSPs} In column 3 (cross-CSP analysis), we observe that different CSPs use different configuration values in NA. While some of the heterogeneity can be attributed to differences in default values across different OSs, we do observe (in column 5) that a nontrivial amount of CSPs are using non-default values, e.g.,
amazon.com using an initial congestion window (ICW) value of 24 MSS.

\textbf{Observation-2: Homogeneity within a CSP:} In column 4 (cross-region analysis), we focus on the configuration used by a specific CSP across different regions. Specifically, this column denotes the percent of CSPs that use a different configuration in NA than in any of the other four regions. We observe that only a small number of CSPs tune their network stack to account for regional differences. 
We observe that the highest amount of tuning occurs at L4, with 7\% of the CSPs tuning the ICW differently in NA than in other regions, e.g., \reword{amazon.de} uses a value of 24 MSS in NA but 10 MSS in Asia.

\begin{figure*}[!t]
	\centering
	\begin{minipage}{0.26\textwidth}
		\includegraphics[width=1.8in]{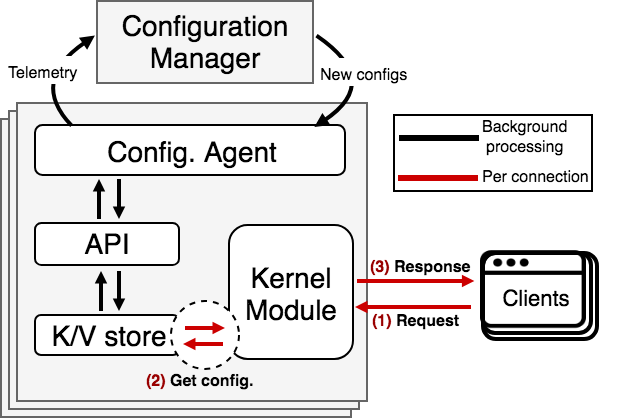}
		\captionsetup{justification=centering}
		\caption{\sysname \\Architecture.}
		\label{f:architecture}
	\end{minipage}
	\begin{minipage}{0.38\textwidth}
		\begin{center}
			\begin{scriptsize}
				\begin{tabular}{p{2.85cm}|p{2.8cm}} 
					\textbf{API Interface}& \textbf{Description} \\ \hline
					ID = initConfigProfile() & Creates new config. profile. \\ \hline
					setProfileParam(ID,key,value) & Sets parameters to profile. \\ \hline
					getProfileParam(ID,key) & Gets parameters for profile.\\ \hline
					deleteConfigProfile(ID) & Uninitializes the profile. \\ \hline
					AssignProfile (SocketID,ID) & Assigns a socket to a specific profile.\\ \hline
					AssignProfile (IP-Prefix,ID) & Assigns a prefix to profile.  
				\end{tabular}
				\captionof{table}{ConfigAgent API}
				\label{t:configagent}
			\end{scriptsize}
		\end{center}
	\end{minipage}
	\begin{minipage}{0.34\textwidth}
		\includegraphics[width=2.3in]{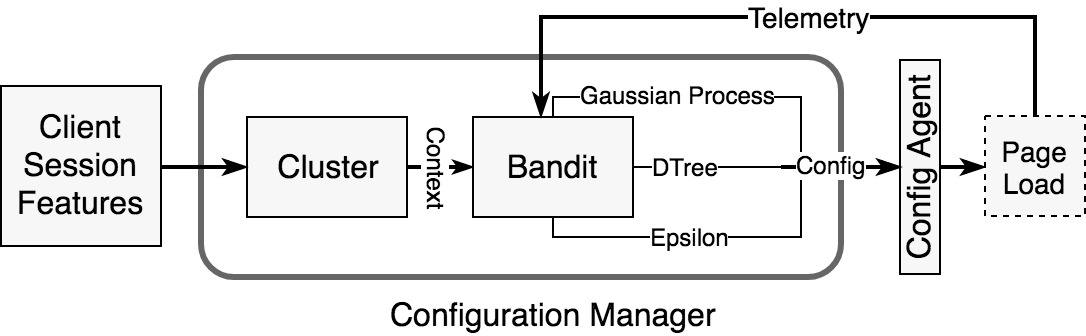}
		\captionsetup{justification=centering}
		\caption{Learning \\Framework.}
		\label{f:learning_framework}
	\end{minipage}
	\vspace{-0.2in}
\end{figure*}

\subsection{Implications of Configuration Tuning}
\label{s:config_study}

Taken together, these observations indicate that while modest tuning is performed on a per-provider basis, this tuning is not specialized to properties of specific regions. Next, we quantify the impact of  reconfiguring the networking stack by conducting a large scale study of the impact of selecting the optimal configurations over the default configurations across different network conditions and websites. 

These experiments are conducted in a testbed: each webpage is loaded five times and the median PLT is computed. Within the testbed, we explored (1) a wide range of realistic network conditions using NetEM -- we extracted the network conditions from four real-world datasets~\cite{fcc,caida,mawi,atc:pantheon} (explained in \S~\ref{s:simulator}), (2) a variety of websites on the Alexa Top-1k list belonging to various categories; news, social networks, sports, business, e-commerce, and entertainment, and (3) the protocol configurations in Table~\ref{t:configurations}. We reconfigured these protocols using IOCTL socket calls, modifying IP tables or modifying application modules.




We begin, in Figure~\ref{f:bestvsdefault}, by exploring the implications of using sub-optimal configurations.
We observe that in the median case, there is a 8-18\% improvement in performance when oracle's recommended optimal configuration is used (found through brute-force exploration in testbed).
While the number may appear small, they can result in tremendous revenue savings in the order of millions~\cite{bing,brutlag2009speed} and more in the developing regions where CSPs are investing heavily to improve those conditions~\cite{fb:drone, google:ballon}. Unsurprisingly, it is within the conditions representative of these developing regions (the tails of the distribution) that reconfiguration provides the most benefits.


\theo{we may wnt to review the next part}

\usama{I have cut the low ICW discussion. Will save space for more interesting stuff.}

\textbf{Dissecting the need for tuning congestion control:}
\label{s:pantheon-data}
Next, we use the dataset from Pantheon~\cite{atc:pantheon} to re-affirm our hypothesis that tuning transport layer is necessary to account for underlying heterogeneity observed in networking infrastructure. Specifically, Pantheon tests different transport protocols from servers spread out across the world. From the Pantheon data-set, we observe that emerging protocols, e.g., BBR, PCC, or Remy, which promise to use probing or machine learning to improve performance, do not provide uniform improvements. In particular, we observed that in many situations BBR is suboptimal, performing 3X to 10X worse than the optimal congestion control. Moreover, no congestion control is optimal for more than 25\% of the networks tested and the median congestion control is optimal for only 6\% of the networks. 
Despite the rapid improvements at the transport layer, there is still a need for configuration tuning and informed-selection of the protocols.

\vspace{-0.1in}
\section{Architecture}
\label{s:archi}

At a high-level, our re-architected edge server consists of four components (Figure~\ref{f:architecture}): 
The \textit{HTTP server} which operates as it does today, serving content and collecting performance metrics for each connection.  %
The \textit{\manager}, runs the learning algorithm (\S~\ref{s:learning}) on telemetry collected from the servers.
%
The \textit{\api} which abstracts vendor specific configuration details and provides a uniform interface for configuring web serving stack's parameters across different vendors and collecting performance metrics.
%
A \textit{\agent} runs on each web server and uses the information received from the \manager to configure the network stack through the \api. 
%





 While this architecture is fundamentally simple, adoption in an incrementally deployable manner is challenged by the following practical limitations:


%
%
\textbf{Interfaces and Abstractions}: Today, a web server's configuration parameters are exposed in an ad-hoc manner and require a combination of \textit{IOCTL} and \textit{setsockopt} commands to tune the kernel. 
Additionally, the appropriate performance metrics are not readily exposed.
%

\textbf{Incremental Deployability:} Many CSPs and CDNs leverage well-established code bases. A key challenge to getting immediate adoption is to employ a solution that exposes the novel interfaces and abstractions, required by \sysname, in a manner that does not require modifying or rewriting existing applications or kernels.

\textbf{Scalable Learning:} Section~\ref{s:learning} presents a data-driven learning framework for identifying and efficiently learning appropriate configurations.
Central to realizing this framework, is a system that scalably supports the modules required to effectively learn appropriate configurations.

\vspace{-0.1in}
\subsection{\sysname-API}
\label{s:sys-api}
The \textit{\api} (Table~\ref{t:configagent}) presents a uniform interface over the web server's serving stack thus abstracts away OS and web server specific details. This simplified interface enables the \textit{\agent} to easily configure the network stack and collect performance metrics without having to understand and reason about vendor specific details or their implications.
%
Central to the design and implementation of the  \textit{\api} is a web server architecture that naturally supports flexibility and fine-grained reconfiguration of individual connections.  Unfortunately, traditional kernels expose only a limited subset of the configuration parameters for flexible reconfiguration. For example, some parameters (e.g. tcp\_auto\_corking) can be configured on the connection level; where as, others can be configured on the process level (e.g. HTTP version) and yet, even others can only be configured on a global scale (e.g., tcp\_low\_latency).  
Using \sysname at a coarser granularity, e.g., global or process-level, either limits the type of connections that can be supported on a machine or limits the space of configurations that \sysname can tune.  To address this challenge, there are several options ranging from user-space TCP/IP stacks~\cite{dunkels2004lwip,jeong2014mtcp,picotcp} and introducing kernel modules or eBPF programs,  to leveraging virtualization (VM/containers).   Below, we briefly discuss challenges associated with different design choices:

\begin{itemize} [topsep=0.1ex,leftmargin=2\labelsep,wide=0pt]
	\setlength\itemsep{0em}
	
	\item Default Linux interfaces (e.g., \textit{sysctl} and \textit{netlink}.):  Most changes made through these interfaces are global w.r.t kernel and all processes and connections share these kernel settings. This provides incremental deployability but limits flexibility.
	\item Containers and VMs: The network stack is shared between all containers and limited in flexibility.
	VM provides the right isolation with the appropriate level of flexibility but the overheads of devoting one VM per configuration are both high and lead to resource fragmentation (\S~\ref{s:microbenchmarks}). 
	
	\item Syscall Interposition: Some socket settings can be changed from user-space through \textit{setsockopt()}. However this requires changing application code which is undesirable. Alternatively, we can employ system call interpositions, i.e., \textit{LD\_Preload}, to intercept system calls and reconfigure the network stack without directly changing the application.
	Although this provides us with the right level of isolation, \textit{LD\_Preload} results in high resource over-heads (\S~\ref{s:microbenchmarks}). 

	\item eBPF: Recent works~\cite{brakmo2017tcp,tran2019beyond} have explored eBPF to tune TCP socket settings. This is a promising direction, in works, and currently lacks tuning all parameters listed in Table~\ref{t:configurations}. 
	
	\item Kernel Module: In face of these limitations, our design of \sysname introduces a kernel module which directly exposes all the parameters and provides direct control over the parameters at the connection level by modifying the kernel's data structures, e.g., SKBuff for Linux. The use of a kernel module ensures that the kernel is not changed while simultaneously allowing us to introduce the functionality required to provide fine-grained control over server's configuration parameters. \theo{how is kernel different from eBPF} \usama{Both follow the same direction and are two design points within kernel space. Both are non-intrusive i-e doesn't make any change to kernel. eBPF is more stable in the sense that it doesn't break/crash kernel.  }
\end{itemize}

%
%
%
%

\theo{Need to add some of this: Moreover, the ConfigAgent collects statistics for each connection (IP address) including the RTTs, loss, bandwidth, and jitter passively and estimates the network conditions faced by the clients.  Note: today, most TCP implementation passively collect these statistics from each client and explose them through well defined interfaces~\cite{web10g}.  The ConfigAgent can extract this information by interacting with these well defined interfaces. The statistical information is aggregated and return to the learning function to minimize overheads.}

\vspace{-0.1in}
\subsection{\agent}
\label{s:agent}
The \textit{\agent} is the glue logic between the \textit{\manager} and \sysname-API --- uses information provided by the \manager to configure the \sysname-API. 
%
%
Our design of the \agent explores the following design choices:



\parab{Push-based design:} Our design chooses a proactive approach, where the \textit{\manager} constantly pushes configuration mappings to the \textit{\agent} which caches them in its local KV. This approach ensures that \sysname does not incur any connection startup delay for waiting for the \textit{\agent} to pull configuration mappings from the \textit{\manager}.


\parab{Imperfect Data:} The \textit{\agent} may not have a configuration mapping, if the incoming client's features (\S~\ref{s:clustering}) are not known (eg. first connection from a previously unseen prefix). In these situations, the \textit{\agent} uses the default configuration and simultaneously queries the \textit{\manager} for a better mapping.

\parab{Long lived Connections:} For long lived connections, the network conditions may necessitate a change in the connection's configurations \footnote{Network conditions change on a time scale of minutes~\cite{zhang2001constancy, balakrishnan1997analyzing, jiang2016cfa}.}. In these situations, the \textit{\agent} receives updates from \textit{\manager} and applies the new configurations to the connection.~\footnote{Recall, \agent periodically sends performance information to \textit{\manager} and receives new configuration mappings based on these performance statistics.}

\vspace{-0.1in}
\subsection{\manager}
The manager runs in a centralized location, e.g., a Data Center or locally in a Point of Presence (PoP). The implications are later explored in \S~\ref{s:cm-design}. It is charged with running the learning algorithms (\S~\ref{s:learning}) and disseminating the configuration maps to the \textit{\agents}' cache (\S~\ref{s:agent}).
The \textit{\manager} disseminates and collects data from the \textit{\agents} using a combination of well-understood techniques. In particular, \sysname leverages a distributed asynchronous messaging system (ZeroMQ~\cite{zeromq}) for all communication between the \textit{\agents} and the \textit{\manager}.  For the configuration maps and the A/B information, \textit{\manager} broadcasts to all \textit{\agents}. Whereas for reporting performance data and for making one-off-request for configuration maps, the \textit{\agents} use unicast.

\usama{I have added some part of discussion wrt pytheus from the google doc to this in a summarized fashion.}
\subsection{Deployment considerations}
\label{s:two-loops}

\newadd{
Prediction workflow operates in two logically separate phases: The \textit{first} involves updating the learning algorithms and Network Class (NC) rules at the Configuration Manager. This is time-consuming process as it involves processing large amount of data. The \textit{second} phase is a fast, real-time process that uses rules generated by first phase for live users.
}

\newadd{
On every model update, the Configuration Manager pushes NC rules (\S\ref{s:clustering}) and learning model decisions (\S\ref{s:learning}) to the front-end servers' Configuration Agents. These rules are translated into key-value stores, eg. learning model decisions are stored as NC (key) and configuration (value). For accurate NC classification, the Agents track real-time network characteristics of the clients (\S\ref{s:clustering}).
Unlike Pytheus~\cite{nsdi:pytheas}, any change in client characteristics triggers a instant change in their NC (and prediction result), as the classification is a simple key-value store lookup, rather any further computation. A network change event (eg. congestion) will result in change in one of the features (eg. congestion at bottleneck link will likely increase RTT), triggering a change in NC as the new features (increased RTTs) might lie in a different NC, according to the NC rules stored at the front-end.
}

	\vspace{-0.1in}
\section{\sysname's algorithm}
\label{s:learning}


\note{removed breadcrumbs}
Improving performance by employing heterogeneous configurations presents an interesting learning problem:
Next, we formulate the problem, identify the challenges, and presents an ensemble to address this challenge.

\textbf{Problem Formulation:} 
Given a set of configurations (\textit{C = $\{c_1, c_2 ... c_n\}$}), network conditions (\textit{N = {$\{n_1, n_2 ... n_n\}$}}), websites  (\textit{W = {$\{w_1, w_2 ... w_n\}$}}) and a function, \textit{f()}, that maps a website, network condition, and configuration to metric of web page performance (e.g., Page Load Time (PLT) or SpeedIndex (ST)).  Note that: \textit{$f(c_i, n_i, w_i)$} returns the web page performance metric value for applying configuration $c_i$  to a client loading website $w_i$ on network $n_i$.  In this paper, we use PLT as the metric for web page performance and note that PLT can be easily replaced with other metrics.
Our goal is to solve equation~\ref{eq:minarg} and find a configuration (\textit{c$*$}) that minimizes \textit{f()} for a given combination of a website (\textit{$w_i$}) and a network condition (\textit{$n_i$}).

\vspace{-0.1in}	
\begin{equation}
\argminA_{c*} f (c*,n_i, w_i) = \{ f (c_i,n_i, w_i) | \forall  c_i \in C \}
\vspace{-0.1in}
\label{eq:minarg}
\end{equation}
\vspace{-0.05in}


Solving the black-box function, f(), requires exploring through the sample space. Two possible candidate algorithms are:
\begin{itemize} [topsep=0.1ex,leftmargin=2\labelsep,wide=0pt]
	\setlength\itemsep{0em}
	\item \textit{Brute force}~\cite{sigcomm:oboe} tests each possible configuration one by one until the entire space has been explored.

	\item \textit{\BO} (BO) \cite{pelikan1999boa, brochu2010tutorial} is an effective global optimization strategy, that uses a prior probability function to capture the relationship between the objective function (f(c,n,w)  Eq\ref{eq:minarg}) and the actual data samples observed. 
	\BO models f(c,n,w) as a Gaussian process \cite{brochu2010tutorial}. The Gaussian process is a distribution of candidate objective functions and is used to select the next promising point, (c*), which is then evaluated on a connection. The Gaussian process then updates its posterior belief by adding the new observation f(c*,n,w) to the set of seen observations. As a result, with each new observation, the space of possible candidate functions gets smaller and the prior gets consolidated with evidence.
	To explore the search space in a principled manner, \BO includes an \textit{Acquisition function} \cite{brochu2010tutorial} which selects the next point to test by calculating the loss associated with computing the next point and selects the point with the least loss as the next target point for testing. 
\end{itemize}


\textbf{Challenges:} Both approaches are sub-optimal in a production setting for our use-case due to several reasons: (1) non-stationary network conditions (network conditions change every few minutes), (2) non-Gaussian noise~\cite{sigcomm:metis} (tail latency can not be modeled as a Gaussian process), (3) BO can be highly sensitive to its hyper-parameters when the underlying data has non-Gaussian loss~(\S~\ref{s:benefits-bandits}), (4) costly data collection (collecting data requires testing on end-users which can impacts PLT and CSP revenue), and (5) data scarcity (testing on individual users requires each user to generate a tremendous number of connections but a user may only visit the site a few times) \newadd{(6) Gaussian process is limited in its ability to model non-continuous space~\cite{sigcomm:metis}}.

 \textbf{Workflow:}
 Figure~\ref{f:learning_framework} presents our process for alternating between a directed and a stochastic search process for building our performance model. Central to the design of \sysname is the intuition during the early phases of modeling-building, the search should be directed to speed the process; however, at steady-state, once a good model is built, the search should be more stochastic to iteratively improve the model and tackle non-systematic noise.

\sysname clusters clients based on similarities and uses these clusters as a ``context'' to determine the quality of the model. Given this context, we employ a multi-armed bandit which alternates between two distinct exploration mode as a function of the ``context''.  For exploiting the search-samples and data, our bandit builds a performance model (i.e., a decision tree) which is used by the exploitation arm to make efficient predictions



\vspace{-0.15in}
\subsection{Learning Optimal Configuration}
\sysname makes two requirements of the learning algorithm: 
First, the algorithm should accurately predict the optimal configuration (\S~\ref{s:dt}). Second, when the algorithm is unable to predict the optimal configuration, it should identify the next samples such that it has a higher probability of improving prediction accuracy (\S~\ref{s:bandits}).

\vspace{-0.1in}
\subsubsection{\textbf{Prediction with Machine Learning}}
\label{s:dt}

%

For prediction, we explored a number of different techniques including Support Vector Machines (SVMs), Decision Trees (D-Trees), and Random Forests (RF).  
%
Using cross-validation, we found their accuracies to be fairly close: SVM 93.3\%, DT 95.6\%,
RF 96.1\%. Decision Tree hits the sweet spot, providing comparable accuracy to other models tested and being efficient enough to build and update at scale. 

A decision tree is a supervised, classification model that predicts class labels for new data items. In our case, a decision tree predicts configuration parameters for new network connections.  Each node, except for the leaves, captures a binary classification decision, predicated on a subset of the feature set. Leaves contain configuration parameters for features which satisfy the classification decision along the path. 

\vspace{-0.1in}
\subsubsection{\textbf{Adaptive Sampling with Bandits}}
\label{s:bandits}


To enable adaptive sampling, \sysname leverages a contextual, multi-armed bandit. Our bandit has three arms: (a) Gaussian process -- to quickly discover a ``good'' solution, (b) epsilon bandit -- to re-sample data points and counter the key problem with the Gaussian bandit, and (c) learning bandit (D-Tree), which exploits knowledge (data samples) gathered by the other bandits to predict effectively. A key distinction between the Gaussian-based exploration and the epsilon-based exploration is that Gaussian-bandit samples based on clusters, whereas the epsilon-bandit samples across the whole population.


\begin{itemize} [topsep=0.1ex,leftmargin=2\labelsep,wide=0pt]
	\setlength\itemsep{0em}
		
	\item \textbf{Exploration Arm-1 (Gaussian process):} The Gaussian process (GP) bandit enables directed exploration to quickly discover a ``good'' (may not be optimal) solution when no information exists for a network class through the use of an acquisition function. 
	There are multiple acquisition functions available \cite{brochu2010tutorial} and we use \textit{Expected Improvement(EI)} because of its well-documented success \mbox{\cite{alipourfard2017cherrypick, gardner2014bayesian, duan2009tuning}}. This search process includes two terminating conditions: a threshold on EI and minimum of number of data points to explore. 
	Gaussian process is better suited than other statistical techniques (coordinated descent, hill climbing, or random sampling) due to its ability to optimize arbitrary black-box functions without domain-knowledge. 
	For example, coordinated descent requires domain knowledge to determine the order inwhich to explore paratemeter dimensions -- barring such domain knowledge, coordinated descent may get stuck in a local optima ~\cite{alipourfard2017cherrypick}.
\note{Some drawbacks removed.}	
	
	\item \textbf{Exploration Arm-2 (Epsilon-bandit):} 
	The Epsilon-bandit (A/B testing) uses a stochastic model to select samples. This allows \sysname to resample old data points and overcome issues endemic with Gaussian process (and Bayesian Optimization in General), e.g., Gaussian noise, non-stationarity, and optimizing parameter dimensions with non-continuous spaces.
	\sout{\newadd{A typical PoP have around 10K connections per minute.}}
	The network operator bounds the random exploration by defining the \textit{degree of randomness}: a parameter that trades-off between the speed and the impact of exploration on end-user QoE. A high degree of randomness improves exploration but results in a negative impact on client's QoE due to constantly changing configurations.
	\usama{Assumption/claim: Random is able to correct the noise and explore the space. We are using the right degree os randomness is another assumption as we are not presenting any results about that.}
	
	\item \textbf{Exploitation Arm}  The exploitation arm uses the prediction algorithm discussed above (\S~\ref{s:dt}). It also aids in bootstrapping a new NC. When a new NC shows up, the exploitation arm predicts a candidate configuration using the trained model. The predicted configuration is used as one of the initial samples in bootstrapping the Gaussian process bandit for the new NC.
\end{itemize}
%

Our ensemble fuses these bandits: in essense, \sysname is constantly switching between exploitation and exploration.  During exploration, \sysname uses the Gaussian Process to quickly discover a ``good'' configuration and subsequently switches to the greedy epsilon-bandit to move towards optimal. The transition from Gaussian to epsilon occurs when a specific parameter (EI) of the Gaussian bandit exceeds a threshold.



\subsection{Discovering Network Classes (NC)}
\label{s:clustering}
\usama{Assumption/claim: I have tried to address the NC related ambiguity by describing different NC techniques that can work with \sysname and then explaining one option that we implemented in our prototype. Also added the offline part of NC building as it doesn't need to be completely online. Also removed features OS, Browser. I think OS and browser doesn't add a lot and makes dimensionality higher. The goal is to address comment about high dimensionality and how much OS/browser can impact TCP.s}

At a high level, \sysname makes the assumption that users with very similar network conditions will require identical configurations.  This assumption is based on studies~\cite{jiang2016cfa,mukerjee2015practical} which show that users with similar conditions observe similar performance.  Thus, \sysname defines \textit{network classes (NC)} as groups of users with similar network properties.  \sysname clusters clients based on properties ranging from performance characteristics (loss, bandwidth, and latency) and device type\sout{(OS, browser, device-type)} to path characteristics (AS-path, client-ISP, geo-location)\sout{(AS-path, City, State, client-ISP)}.   
\footnote{This data is collected throughout the connection's life time and as more data is collected the connection's cluster is re-evaluated. During the TCP handshake, the path and performance characteristics are collected using IP address and publicly available data (e.g. RouteViews). The performance characteristics are estimated using the packets exchanged during the handshake and refined as more packets are exchanged.
The device characteristics are captured once the user-agent string is captured.}  

Clustering can be automatic, using conventional techniques, e.g., K-means or hierarchical clustering, and by using domain-knowledge, e.g., Facebook network class~\cite{fbconn}, PolicyAtoms~\cite{caida:policyatoms} Hobbit~\cite{imc:hobbit} or CFA~\cite{jiang2016cfa} .
\newadd{
Alternatively, CFA~\cite{jiang2016cfa} can be used to reduce feature set to only those features that matter -- tackling the potential challenge of data sparsity due to high dimensionality.
Although \sysname can incorporate any of the aforementioned techniques, our prototype uses kmeans to classify clients in network classes. As the objective is to group clients with similar web performance, the number of classes (or k) is determined by evaluating the spread of performance metric within a NC (bounded by one standard deviation from mean). We use kmeans due to its simplicity. \sysname initializes with offline-generated NCs (using historical traces and testbed simulations) that evolve with time if performance of clients within same cluster diverge. We do not implement CFA due to unavailability of public dataset with required features (\S\ref{s:simulator}).
}
\usama{Assumption/claim: Is the selection of k to bound the spread of performance within NC enough?}






\note{removed bootstrapping direction}
\subsection{Design Choices for Gaussian Process}
\label{s:bootstrap}
\usama{Added the alternate boot strapping through offline exploration. Replacing the old bootstrapping with offline can have cascading effects on eval, as some experiments use it for explaining the trends. I think we should mention it as an alternate option or add this to discussion.}

In addition to the acquisition function, and stopping parameters, designing a gaussian process requires choosing a bootstrapping method. In our situation, the bootstrap methodology is especially crucial for ensuring that the Gaussain-Bandit quickly finds appropriate samples.
%
Recent works~\cite{alipourfard2017cherrypick, duan2009tuning, Bilal:2017:TAP:3127479.3127492} have demonstrated the applicability of three distinct bootstrapping approaches: (i) \textit{random}, in which the initial configurations are randomly selected; (ii) \textit{domain-specific}, in which prior domain knowledge, captured through operator interviews or simulations offline, are used to rank configurations to sample;
(iii) \textit{Latin Hypercube Sampling} (LHC) which divides input space into partitions and selects a sample from each partition to spread the samples evenly across space~\cite{lhs}.

In this work, we use LHC to bootstrap the learning process.
LHC has been found to aid bootstrapping \BO by reaching an optimal decision quicker \cite{mckay1979comparison}.
We observed LHC to speed up exploration in comparison with others by reducing the number of optimization steps by 2-3X,  as the bootstrapping samples are spread evenly across space.
A perfect rankings of configurations cannot be known prior to actually testing configurations, leading to ranking-based bootstrapping being sub-optimal to LHC.
In addition to LHC, \sysname's exploitation arm also aids in selecting the bootstrapping samples for new network classes (\S~\ref{s:bandits})


\usama{Added the alternate boot strapping through offline exploration. Replacing the old bootstrapping with offline can have cascading effects on eval, as some experiments use it for explaining the trends. I think we should mention it as an alternate option or add this to discussion. For the time being, I have striked-through the newlyy added text about offline bootstrapping.}

	\vspace{-0.1in}
\section{Prototype}
\label{s:prototype}
In this section, we describe our prototype of \sysname, focusing on the implementation highlights.




\noindent \textbf{\api:} is a user-space application that updates settings in kernel via a kernel module exposed as a character device~\cite{char-device}. Our use of a kernel module to tune the network stack's enables us to deploy our system without changing the kernel.
\note{removed the KV state.}
We used Linux's vanilla implementation for key-value store \cite{kernel-hashtable}.

\noindent \textbf{\agent:} has an agent residing within kernel module and gets its cues from \api to tune configurations. It wraps functions provided by kernel's congestion controls through \textit{tcp\_congestion\_ops} and reuses their functionality, while at the same time changing appropriate fields in \textit{sock} and \textit{dst\_entry} structs. For reporting a connection's performance stats, its user-space part logs stats from a socket's \textit{tcp\_info} struct and Apache's logs. HTTP is tuned by routing requests to differently tuned apache instances.

\noindent \textbf{\manager:} is developed in 1435 LoC (Python). 
The \manager uses SciLearn~\cite{sklearn-dt} for D-Tree and GPyOpt~\cite{gpyopt} for \BO. 
For communication between the \manager and \agents, \sysname uses ZeroMQ~\cite{zeromq}. 
For D-Tree, we use an optimized version of CART algorithm (implemented in SciLearn by default).
We use entropy for the information gain as the measure of quality of split, 80 as the minimum number of leaf nodes, 2 as minimum number of samples needed for split and do not limit depth of tree.  For Gaussian process, we use init\_sample=4, min\_sample\_tested=7 and EI=5\% thresholds.
	\vspace{-0.15in}
\section{Evaluation}
\label{s:eval}
To understand and quantify the benefits of reconfiguring the networking stack, we evaluated \sysname using both a large-scale trace driven network simulator and a live prototype deployment of around hundred clients. Together, these two approaches enable us to understand the behavior of \sysname under real and dynamic conditions as well as to simultaneously analyze the implications of individual design choices on the entire systems.


\vspace{-0.15in}
\subsection{Large Scale Trace Driven Simulations} 
\label{s:simulator}

\begin{figure*}[tb!]
	\centering
	\begin{minipage}{0.45\textwidth}
		\centering
	\includegraphics[width=2.8in]{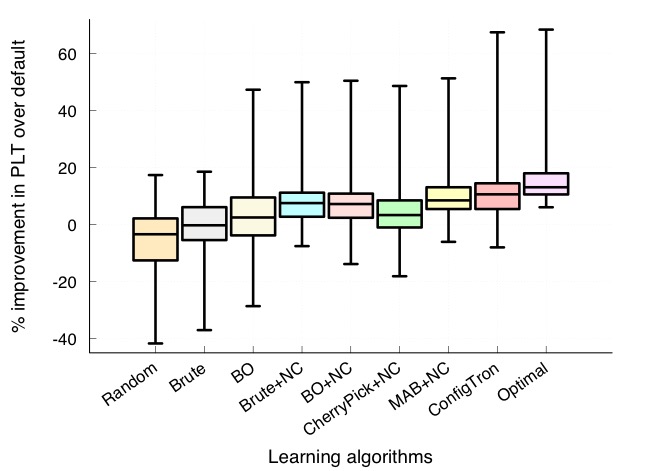} 
		\vspace{-0.1in}
		\caption{CAIDA traces}
		\label{f:caida_bf}
	\end{minipage}
	\begin{minipage}{0.45\textwidth}
		\centering
		\includegraphics[width=2.8in]{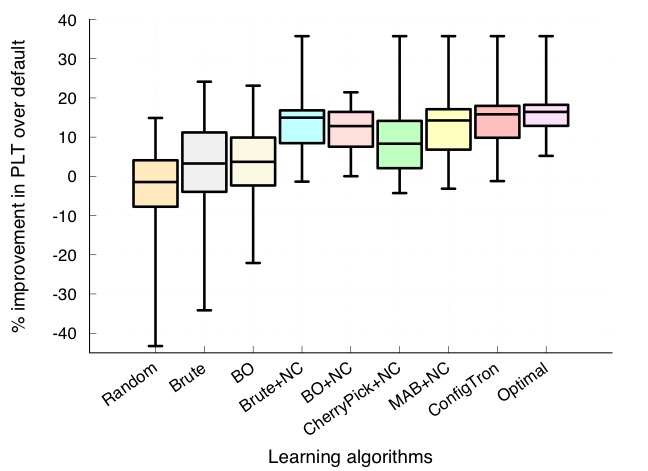} 
		\vspace{-0.1in}
		\caption{MAWI traces}
		\label{f:mawi_bf}
	\end{minipage}
	\label{f:sim-results}
	\vspace{-0.2in}
\end{figure*}

\usama{Assumption/claim: The network conditions, extracted from traces, capture all the features discussed in the section 3. Perhaps we should have some sort of a disclaimer.}

\textbf{Datasets:}
To simulate client activity, we use data from four sources: 
(i) \textit{CAIDA}~\cite{caida}, packet traces collected at Equinix data-center in Chicago in 2016, 
(ii) \textit{MAWI}~\cite{mawi}, packet traces from Japan WIDE backbone (in December 2017), 
(iii) \textit{FCC}~\cite{fcc}, a nation-wide home broadband dataset collected by the FCC,
(iv) \textit{Pantheon}~\cite{atc:pantheon}, a data set of client sessions 
between different regions.
This work does not raise ethical concerns as the public datasets used are anonymized.

\newadd{
\textbf{Generating client sessions:}
A client session is defined by 3 metrics: (a) session characteristics (bandwidth, latency, loss rates) at the start of connection, (b) time duration after which network changes, (c) sessions characteristics after change.
Our simulator processes two categories of client's session traces:
(i) \textit{Real-world client sessions} are extracted from CAIDA and MAWI datasets. Using the packet traces from these datasets, session characteristics between a pair of end-points and any temporal changes are calculated, 
(ii) \textit{Simulated client sessions} use bandwidth, latency and loss rates distributions, extracted from FCC and Pantheon datasets. Changes in network (duration after which network changes and revised session characteristics) follow distributions from real-world packet traces (CAIDA, MAWI). For detecting and measuring change in network, we used Bayesian Online Changepoint Detection algorithm~\cite{adams2007bayesian} due to its effectiveness for similar use-case~\cite{sigcomm:oboe}. An IP-distribution, extracted from CAIDA, MAWI traces, models the temporal aspect of client's connections (time at which a client connection (or IP) is seen in trace).}
Collectively, with the four datasets we are able to simulate $\sim$16M sessions 
from different geographic regions thus providing us with good coverage over a wide range of representative network conditions.


\textbf{Optimal Configurations:} Using a testbed, we re-create the network conditions from traces and for each network condition we emulate all combinations of configurations (Table~\ref{t:configurations}). 
For each \{network condition, configuration\} pair, each webpage in our corpus is loaded multiple times with a browser and the performance, or page load time (PLT), of each \{configuration, website\} is measured.
The final result is stored in a large tensor that maps \{network condition, configuration, website\} to PLT -- called the \emph{PLT-Tensor}.

\usama{Assumption/claim: Simulator is capable to capture the TCP dynamics perfectly, although we don't have any competing, background flows.}

\textbf{Simulator (Virtual Browser):}
Our simulator takes the PLT-Tensor and client session traces as input.
The simulator processes the trace and simulates the client's browsing behavior and interaction with \sysname by (i) processing the time series to characterize the connection~\footnote{Each end-user is modeled as a time series -- measured network conditions indexed over time. This time series enable us to simulate client sessions by modeling the changes in network conditions, faced by these clients.}, (ii) using the learning framework in \S\ref{s:learning} to determine the  appropriate configuration to apply to this connection, (iii) simulating the webpage load process by using the PLT-Tensor to determine PLT for the client given the selected configuration, and (iv) feeding the result of the simulation back into configuration manager to enable continuous learning.
%



\noindent\textbf{Limitations:}
\sout{
To ensure precise control over the parameters being analyzed, we do not generate excessive load on the client or server and do not re-use TCP connections. As a consequence, we are unable to analyze a subset of TCP parameters, e.g., slow\_start\_after\_idle which requires re-using TCP connection after idle time.
As part of future work, we plan to explore systematic approaches to incorporate these parameters.}
The PLT-Tensor does not include \textit{TCP slow start after idle} as we do not re-use TCP connections in testbed to ensure clean measurements.
Additionally, the simulator is unable to emulate different end-user devices.

\textbf{Alternate algorithms}
We compare \sysname against:

	
\noindent \textbf{i-\textit{Brute-force (Brute):}} For each client, this algorithm exhaustively explores all configurations -- a distinct configuration per client-connection --  and use the best one. 

\noindent \textbf{ii-\textit{Brute-force w. Network Classes (Brute+NC):}} Brute force where exploration is spread out across each network class. 

\noindent \textbf{iii-\textit{Bayesian Optimization (BO):}} Bayesian Optimization for individual user.

\noindent \textbf{iv-\textit{Bayesian Optimization w. Network Classes (BO+NC):}} Bayesian Optimization across groups of users.
	
\noindent \textbf{v-\textit{CherryPick with Network Classes (CherryPick+NC):}} BO+NC with hyper-parameters specified in~\cite{alipourfard2017cherrypick}.


\noindent \textbf{vi-\textit{Multi-armed Bandit with Network Classes (MAB+NC):}} uses multi-armed bandit algorithm with different configurations as the arms, using weighted epsilon-greedy agent. This strategy uses one arm per config. and build individual models for each NC.
	
\noindent \textbf{vii-\textit{Optimal:}} Puts an upper bound on improvement. Optimal parameters for a session are determined by searching the PLT-Tensor for the configuration that minimizes PLT. 
		




Note: \sysname maintains an estimate of a client's network using its historical interactions. In case when client's network changes, optimal uses the the new, correct view of network, whereas \sysname and other baseline algorithms use the historical estimate for prediction.

\vspace{-0.15in}
\subsection{Effectiveness of \sysname}
\label{s:benefits-sim}

Figures \ref{f:caida_bf},\ref{f:mawi_bf} present the improvement in PLTs over default configurations (Table~\ref{t:configurations}) across the different algorithms for multiple traces. 
\sysname outperforms all alternatives, consistently delivering lower PLTs. \sysname improves median PLTs by 19\% for Pantheon (500ms), 16\% (750ms) for MAWI, 11.2\% (280ms) for FCC and 10.1\% (250ms) for CAIDA. 

Unlike \sysname, the static strategy (Default) is unable to account for differences between networks and websites (\S \ref{s:motivation}).
While, the dynamic strategies (Brute and BO) are able to apply different configurations to different users, these approaches assume that the network remains static and are unable to adapt to fluctuations in network behavior. Moreover, due to its inability to adjust to fluctuations, BO often explores over 90\% of configuration space without achieving the target EI. Brute+NC, BO+NC and CherryPick+NC improves over the prior by amortizing the costs of learning but fail to adjust to non-Gaussian variations.
\sout{Effectively improving PLT requires configuration-tuning that adjusts to both differences in networks and variations of network conditions overtime. The precludes a large body of systems-tuning techniques~\cite{nsdi:earnest,van2017automatic,alipourfard2017cherrypick}, which do not effectively adapt to both sources of variations. CherryPick's EI threshold (10\%) stops exploration pre-maturely and NCs get stuck at local optimal. Errors in EI estimation for BO are due to imperfect modeling for the discrete features and noise (\S\ref{s:learning}).}

\theo{there are several trends here that we can discuss: (1) the tails are different. (2) in CAIDA there are distinct patterns, (3) Pantheon the patterns are more in the median and lower tail. butwhy do they have almost identical upper tail?  (4) why is the upper tail for MAWI so interesting?}

\newadd{
We observe fairly similar improvements for FCC and CAIDA as both traces cover network conditions from USA. The improvement over default increases for MAWI and Pantheon traces, as they have higher RTTs and packet loss rates. Eg, their 75th percentile RTTs are greater than 100ms, as compared to $~$34ms for FCC and CAIDA. We observe that higher reconfiguration benefits are seen for network conditions representative of under-developed regions (high RTTs and loss rates, low bandwidths), as the default configurations are mostly tuned for developed regions networks~\cite{dukkipati2010argument, wang2014speedy}. 
}


%

\vspace{-0.1in}
\subsection{Benefits of Contextual Bandits} 
\label{s:benefits-bandits}

\begin{figure}[b!]
	\vspace{-0.2in}
	\centering
	\begin{subfigure}[b]{0.23\textwidth}
		\includegraphics[width=1.5in]{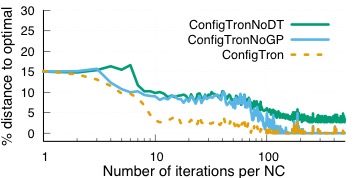} 
		\caption{Cold start}
		\label{f:convergence-cold}
	\end{subfigure}
	\begin{subfigure}[b]{0.23\textwidth}
		\includegraphics[width=1.5in]{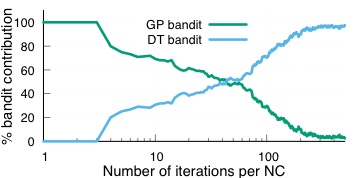} 
		\caption{Bandits contribution}
		\label{f:bandit-contrib}
	\end{subfigure}
	\vspace{-0.1in}
	\caption{Convergence}
	\label{f:convergence-results}
\end{figure}

Next, we analyze the benefits of \sysname's multi-armed bandits. We analyze two versions: \sysname-NoGP, a version of \sysname without guided exploration.
, and \sysname-NoDT, a version of \sysname without the decision tree.
For the two alternatives, Figures~\ref{f:convergence-cold}, plots the median distance from optimal across all network classes, for a given iteration of algorithm optimization.

The key observations are that: (i) initially all algorithms perform comparable because of a lack of data (iterations 1 fig\ref{f:convergence-results}), (i) as more data is gather \sysname outperforms both approaches because of its ability to blend benefits of both approaches -- essentially efficient exploration and effective exploitation (iterations 3-10), (iii) eventually, with sufficient data \sysname-NoGP is able to use the decision tree's predictive power to achieve near ideal performance (iterations 100+), 
finally (iv) while \sysname and \sysname-NoGP perform comparable for median performance at steady-state when sufficient data exist, performance at the tail is \textbf{still} different -- the use of guided exploration enables \sysname to explore specific configurations parameters that improve tail latency.
\note{Removed discussion of why BO is ineffective -- this is redundant and should be already mentioned above}

\theo{we should be abel to discuss tail performance using figure 4,5,6.}
\theo{this seems redundant and should be integrated above}
Finally, In Figure~\ref{f:bandit-contrib} we analyze the contribution by each bandit. Initially when \sysname is exploring new configurations, GP bandit is largely used for a guided exploration. However as more data is collected with time, GP bandit's contribution is overwhelmed by DT bandit, highlighting the benefits of greedy exploitation. This result stresses the point that a per-class guided exploration is over-shadowed by across-class exploitation, when large data is available.
\theo{what is this cross overpoint}
\textbf{Takeaways:} The complexity provided by \sysname's contextual bandit enables \sysname to simultaneously provide good median and excellent tail performance. 
\subsection{Design Choices}


Next, we analyze the impact of several of \sysname's design choices on the performance of \sysname. 

\obj{(i) NC classification should be on (user,CDN server) pair level. For a big CDN, a single client may classify into different classes for different CDN servers, eg, client C to CDN server 1 latency is 20ms, C to CDN server 2 latency is 50ms. Then it might be sorted into different NCs. (ii) What will be the number of NCs in a real setting? It will be some sort of combination on number of network conditions, device types etc.}

\subsubsection{\textbf{Configuration Manager Design}}
\label{s:cm-design}

\obj{how does latency between learnign agent and config manager/agent impact the performance?}
\obj{Ran a new experiemnt here. This one was for a subset of trace and only focuses at the start of learnign. Things should be pretty different at later stages. }
The \manager can run locally in a PoP or more centrally within a data-center. The choice of location is a fundamental trade-off between the size of data available for learning and the speed with which the system can react to changes: 
An appropriate choice ensures high accuracy while providing a delicate balance between both extremes.

We evaluate both scenarios in our simulator: 
In the local design, there's a separate \manager for each PoP (trace), while for the global case there's a single \manager which for all traces (PoP).
We evaluate several situations, with different latencies between \managers and the web servers.
\note{removed figure}
We observe that while the global \manager is able to make slightly better predictions at the tail (less than 2\% relative)
In particular, we observed that despite the larger data set, the global \manager does not show much improvement in predictions because the different traces are for different regions with distinctly different network conditions (only 17\% overlap). 
%
%


\usama{Assumption/claim: We are simulating multiple days from trace that doesn't capture whole days.}

\subsubsection{\textbf{Frequency of model updates}}
\label{s:update-freq}

The frequency of model updates has a significant impact on \sysname performance -- while existing work have demonstrated that updates as infrequent as hour may suffice~\cite{nsdi:pytheas, jiang2016cfa}, our discussion indicate that operators may want to explore less frequent updates.  Next, we analyze the impact of updating our performance model less frequently: we explore a range of values from every 2 minutes~\footnote{ It takes $\sim$2 minutes to update the models for 10K sessions.} up to every day.  We observed that performance at the tails degraded whereas performance at the median remated relatively stable.

\subsubsection{\textbf{Microbenchmarks}}
\label{s:microbenchmarks}
Next, we analyze the impact of the \agent on the web servers. Specifically, we explore CPU, memory and latency overheads of \agent and the performance implications of having \agent interpose on each connection.  We conducted these tests using the \AB tool~\cite{ab}: we conduct our experiments using two servers. One to run the \AB tool and another to run the \sysname-enabled server. We repeated tests in multiple network conditions, with multiple number of concurrent connections and for multiple websites.
We compared \sysname against a vanilla server using \textit{Cubic} (default setting). 
\AB tool showed no overhead of using \sysname's kernel module in term of latency. Similarly, we observed CPU and RAM overheads less than 1\%. Any RAM overhead is attributed to kernel module's implementation of the KV-store \cite{kernel:hashtable} used for storing the IP masks to configuration rules.

\textbf{Alternative Design Choices:} We also evaluated alternate design choices described in \S~\ref{s:microbenchmarks}.
For VMs, we used one VM for each configuration and used Open vSwitch (OVS) \cite{ovs} for routing flows to the appropriately configured VM.  
We also explored the use of \textit{LD\_Preload} to intercept system call and tuned socket using \textit{setsockopt()}. 
In comparing both choices with \sysname, we observed that the VM-based approach introduced a ~20\% increase in latency where as the \textit{LD\_Preload} introduced a much smaller latency of 2.2\%. We also observed overheads for CPU and Memory utilization: the VM-based approach introduced ~30\% while the \textit{LD\_Preload} introduced a ~5\% increase.

\theo{there's no discussion of the container overheads? is it possible to get some eBPF numbers by using the paper/project from UCL?}

\textbf{Takeaway} \sysname introduces minimal overheads on the webserver (less than 1\%) and is competitively better than alternative design choices which introduce as much as 30\%.

\subsection{Fairness Implications}
\label{s:fairness}


\begin{figure}[tb!]
	\begin{center}
		\includegraphics[width=2.4in]{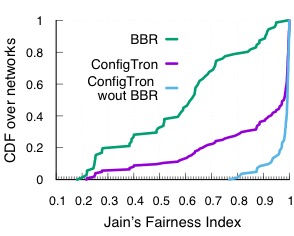} 
		\caption{Jain's Fairness}
		\label{f:jains}
	\end{center}
\end{figure}

As \sysname configuration space covers regular configurations used by linux kernel and current CDNs~(Table~\ref{t:ig}), fundamentally, its fairness implications are bounded by the configurations used in-the-wild today. A number of CDNs use BBR (found to be unfair to Cubic~\cite{ruth2019empirical, hock2017experimental, turkovic3interactions}) and high ICWs (eg. 30 as ICW).
In order to evaluate fairness, we simulate an end-user's last-mile connection in our testbed. We test multiple representative network conditions (3G, 4G, Cable, DSL etc) with shallow buffers. 
In each network, we start 5 \textit{background} flows with default Cubic configuration. 
We further inject a \textit{test} flow matching either (i) \textit{\sysname} (each of \sysname's top 10 configuration) or (ii) \textit{default BBR}. 
Modern browsers commonly spawn 6 connections for HTTP/1.1 and these experiments evaluate config. fairness among the connections.
We collect per flow statistics (throughput, RTTs, loss rtaes etc.) using \textit{TCP\_Info} from server. We further compute \textit{Jain's fairness index} to measure the throughput fairness at steady state.

Figure~\ref{f:jains} presents the results.
Similar to recent studies~\cite{ruth2019empirical, hock2017experimental, turkovic3interactions}, we found BBR to be mostly unfair to Cubic. However, \sysname is fair to Cubic (Jains's $>$ 0.95) for more than 50\% of the cases. 
\sysname's top configuration include Cubic, Vegas and BBR with ICW ranging from 10-30. We do not observe ICWs to significantly impact fairness at steady state as different ICWs with similar CC were observed to be fair with each other. We observed the use of Cubic and Vegas in 6 of top configurations to contribute to \sysname's good Jain's fairness index.
4 BBR configs in \sysname's top 10 mostly  contribute to the unfairness. Removing these configurations (ConfigTron wout BBR) makes \sysname even more fair to Cubic for more than 90\% of the cases.

\subsection{Critical Features and Parameters}
\label{s:dimensionality}

\begin{figure}[htb!]
	\centering
	\begin{minipage}{0.22\textwidth}
		\centering
		\includegraphics[width=1.5in]{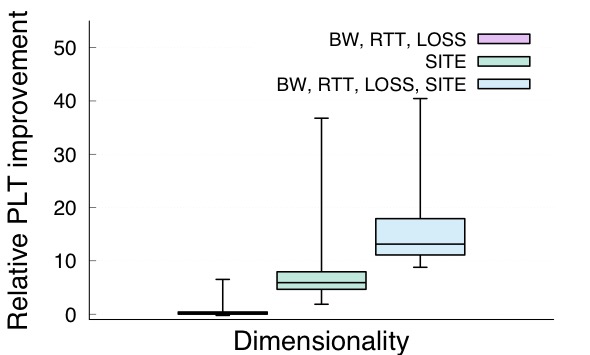} 
		\captionsetup{justification=centering}
		\vspace{-0.1in}
		\caption{Critical \\features}
		\label{f:dimensionality}
	\end{minipage}
	\begin{minipage}{0.25\textwidth}
		\centering
		\includegraphics[width=1.7in]{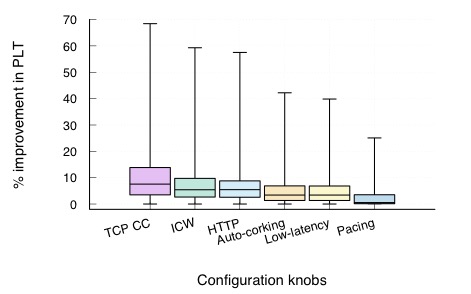} 
		\captionsetup{justification=centering}
		\vspace{-0.2in}
		\caption{Critical \\configuration parameters}
		\label{f:critical-para}
	\end{minipage}
	\vspace{-0.2in}
\end{figure}

In order to understand the impact of input features and configuration parameters on prediction quality, we modified our simulator to train on different subsets of the feature set (or configurations). In particular we focused on subsets of the following features: \{website, throughput, latency, packet loss\} and configuration knobs: \{Table~\ref{t:configurations}\}.

\textbf{Input Features:} 
Recent work~\cite{pi2018ap} suggests latency as good metric for client aggregation. Using latency as a baseline, Figure \ref{f:dimensionality} plots relative improvement when a set of features is used.
The results show that web complexity is a crucial parameter. Without including web-complexity in the feature-list (red-line), median performance is unchanged. This reflects the fact that the impact of network conditions vary depending on the properties of the website -- a long understood insight~\cite{nsdi:klotski, wang2014speedy, netravali2016polaris, nsdi:wprof}.  Upon further analysis, we observe that impact of web-complexity varies. For example, including web-complexity of simple websites, e.g., Search, provides little benefits where as introducing web-complexity for content-rich sites, e.g., shopping, provides significant benefits to learning.

\theo{This figures may be better if it were relative. and there are 3 patterns -- (1) site, (2) everyone, (3) different network combination. Why does Site have a better tail performance than optimal?}
\usama{I checked the data. Site being better at tail was due to gnuplot filtering for outliers (using the IQR range).}

Next, we analyze the relative importance of reconfiguring different configuration parameters. Our goal is to understand the crucial parameters that must be tuned to significantly improve performance. 

\textbf{Configuration Knobs:} Figure~\ref{f:critical-para} plots the results for five configuration options. 
\theo{we need a better figure here.}
%
There are several crucial observations: while the HTTP parameter is crucial. It is not the most important and it is infact third after TCP-CC and ICW.
Unsuprisingly the congestion control algorithm (TCP CC) is the most important parameter because this directly controls throughput. The next important, ICW, is a related parameter that aids the congestion control algorithm -- and is crucial for short flows.  
While there are many other parameters, e.g., Auto Corking, in the median case these provide little benefits but provide tremendous benefits at the tail. This implies that improving tail performance requires turning a broader range of parameters to account of the edge conditions that exist at the tail.








\vspace{-0.15in}
\subsection{Live Deployment}
\label{s:live}
We deployed \sysname on several AWS servers each with 4 cores Intel(R) Xeon(R) CPU E5-2676 v3 @ 2.40GHz and 16 GB of RAM. One of the servers runs the \manager. We evenly divide the remaining servers: half running \sysname-enhanced web servers and the other half running traditional web servers.
%
Our clients were distributed all across the globe using Speedchecker~\cite{speedchecker}. These web clients periodically conduct back to back pageloads for websites from both the \sysname-web servers and the traditional web servers.
We tested a variety of websites, including websites from Alexa top-100.
In total, we had approximately 3200 clients spread across 6 of the continents. The clients conducted pageloads throughout the day, resulting in $~$100K pageloads in 10 days.





\begin{figure}[tb!]
	\begin{center}
		\includegraphics[width=2.4in]{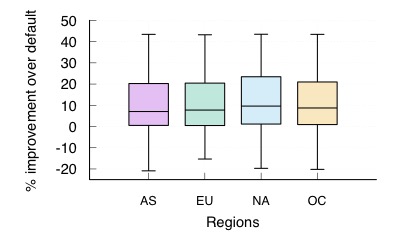} 
		\vspace{-0.1in}
		\caption{Live deployment}
		\vspace{-0.3in}
		\label{f:ld}
	\end{center}
\end{figure}

We observed that \sysname improves the web performance by 8-10\% in the median and $~$43\% in the 95th percentile case.
\usama{I am adding details about regions and best configs. Unfortunately the ditributions of number of pageloads is skewed towards EU and NA. Speedtest has daily as well as hourly limits on clients. As the number of their active clients are much smaller in regions like AF, the number of pageloads there are smaller than EU or NA (EU NA has more than 4X pageloads than NA).}
For the highest improvements scenarios, the optimal configuration include either BBR or Vegas for congestion control.
We observe a stark difference in the ICW values for optimal configurations: clients in developed regions (EU, NA) use much higher ICW (30-50 MSS), whereas under-developed regions employed comparatively lower ICW (16-24). Note that, optimal ICW for both regions is different from default (10 MSS).
Unsuprisingly, majority of the websites which experience greater than 20\% improvement are content-rich, eg. sports, news etc.

	\section{Discussion and Future Work}
Here we discuss open questions in \sysname's design:


%

\parab{Offline exploration:} Recent works~\cite{sigcomm:oboe} have highlighted the effectiveness of offline exploration. While the effectiveness of offline exploration is limited for high-dimensionality situations~\cite{jiang2016cfa,nsdi:pytheas}, we believe that offline exploration can enhance our approach by providing a pre-build model (or warm model) thus reducing model convergence or providing data for proactively pruning in-effective configurations knobs~\cite{van2017automatic}. As part of future work, we are extending our learning model to incorporate pre-build models.

\parab{QUIC, BBR:}  While \sysname has focused on more traditional stacks which use TCP, \sysname can operate over new transport layers, e.g., QUIC or BBR. In fact, both QUIC and BBR introduce a larger number of configuration options, in turns, strengthens the case for self-tuning like \sysname.

\parab{Client Side Changes:} Currently, \sysname requires no client side changes. However, client side changes can improve \sysname's performance in multiple ways. 
As part of future work, we are exploring methods to use emerging Javascript APIs (e.g., NetworkAPI~\cite{networkAPI}) to gather client-side features and incorporate them into our learning algorithm.

\parab{Security and Equilibrium:} Potential implications of self-learning systems include adversarial attacks~\cite{sun2015raptor} or oscillations. We are working to formulate the interactions between different instances of \sysname (i.e., deployments by different CDNs) as a game-theoretic problem to understand the behavior of our system at equilibrium.

\note{removed operational complexity -- peak v. non-peak configurations.}
\sout{\parab{Operational complexity:}
Discussion with major CDN operators have highlighted that they have TCP tuning layers in the networking architecture, eg. Fastly uses VCL to tune TCP on connection level~\cite{vcl-fastly}, Facebook's TCP-BPF~\cite{brakmo2017tcp}. Thereby integrating \sysname into current infrastructure would not require a complete over-haul of edge, rather incremental changes.}

\parab{Management Overheads:} Dynamically reconfiguring the CDN's protocol stack complicates performance diagnosis and troubleshooting. We plan to investigate methods for reducing this complexity, e.g., minimizing the number of active configuration combinations.

\sout{ \parab{Reconfiguration at Higher Layers:} We envision that \sysname will be extended to control higher layers of stack, e.g., to configure different compression algorithms~\cite{singh2015flexiweb}, multimedia-resolution sizes~\cite{anand2016enhancing}, and HTTP scheduling practices~\cite{ruamviboonsuk2017vroom} based on the network's properties and the web site's complexity.  As part of future work, we are working to extend the range of our configuration knobs to include these application level parameters.
}
\textbf{Broader Evaluations and QoE Metrics:} As part of ongoing work, we are planning to understand the limits of \sysname by evaluating \sysname across a broader set of web page QoE metrics (e.g. SpeedIndex).





	\section{Related Work}

\begin{table}[tb!]
	\begin{center}
		\begin{scriptsize}
			\begin{tabular}{ccccc} 
				& \textbf{Online} & \textbf{Global } & \textbf{Cross} & \textbf{Exploration} \\ 
					& \textbf{Data} & \textbf{ Data} & \textbf{Layer} &  \\ \hline
			\sysname & \checkmark & \checkmark & \checkmark & \checkmark  \\ \hline
			Remy~\cite{winstein2013tcp} & X & \checkmark & X & X \\ \hline
			PCC~\cite{dong2015pcc} & \checkmark & X & X & \checkmark  \\ \hline
			Pytheus~\cite{nsdi:pytheas} & \checkmark & \checkmark & X & \checkmark  \\ \hline
			Custard~\cite{jay2018internet} & \checkmark & X & X & \checkmark \\ \hline
			QTCP~\cite{li2018qtcp} & \checkmark & X & X & \checkmark \\ \hline
			Cloud tuning~\cite{alipourfard2017cherrypick, sigcomm:metis, zhu2017bestconfig, Bilal:2017:TAP:3127479.3127492} & \checkmark & \checkmark & X & \checkmark \\ \hline
			DB tuning~\cite{duan2009tuning, van2017automatic} & X & \checkmark & X & \checkmark
			\end{tabular}
			\caption{A comparison of algorithms. \theo{I replaced Autotune with cross layer. We should only include autotuned here. we should have a row for the DB papers, and another for the cloud papers.  Not sure if probing is discussed below -- a better word could be Exploration.}}
		\end{scriptsize}
	\end{center}
\vspace{-0.4in}
\end{table}







\textbf{Web Performance} Many measurement studies~\cite{wang2014speedy, erman2015towards, ghobadi2012rethinking, al2011overclocking, dukkipati2010argument} have explored the performance of different networking protocol settings and the impact of tuning on web performance.  
Our system builds on the observations from these studies: namely that different configurations are required for different network conditions and websites. 

\theo{this cover cross layer. why isn't the RL work discussed here?}
\usama{I had RL stuff at the end of this subsection -- that got commented out.}
\textbf{ML + Transport} Existing applications of ML to the transport protocol~\cite{winstein2013tcp, schapira2017congestion, dong2015pcc,jay2018internet, li2018qtcp} focus on tuning specific aspects of stack and thus provide limited benefits relative to \sysname which tunes across a broader set of layers and parameters(see Section~\ref{s:motivation}). Similarly, while these techniques~\cite{dong2015pcc} use features from the network, \sysname incorporates application features (e.g., website complexity).

Unlike~\cite{winstein2013tcp} which rely on priori assumptions of the network, \sysname builds a performance model-based on live feedback which allows it to adapt to network dynamics. 
%
Where as others~\cite{jay2018internet, li2018qtcp,dong2015pcc} build models using real time data, they operate on a different granularity -- controlling the sending rate of individual servers based on local data (lacking a multi-session view). Instead, \sysname operates at the broader level -- using data from multiple connections to build a model that controls protocol configurations and parameters, thus making it immediately deployable. 

In contrast with~\cite{balakrishnan1999integrated}, \sysname builds on contrasting motivation that different CC strategies (eg, loss, delay, bottleneck-bandwidth based etc) react differently to distinct network paths, whereas CM uses same CC (ECN and loss-based, AIMD) for the divergent paths.
Unlike \sysname which focuses on control over server configurations, others~\cite{ruffy2018iroko,ghobadi2012rethinking} require control over network and endhost to perform appropriate learning and tuning --- applicable to data centers.

\textbf{Self-tuning Systems:} 
The closest related work attempt to tune the configurations of Map-reduce~\cite{herodotou2011starfish}, Databases~\cite{duan2009tuning, van2017automatic}, TCP initial congestion window~\cite{tcpwise, niereducing}, cloud computing~\cite{alipourfard2017cherrypick, sigcomm:metis, zhu2017bestconfig, Bilal:2017:TAP:3127479.3127492} and video~\cite{mao2017neural, nsdi:pytheas, sigcomm:cs2p, sigcomm:oboe} to improve application-specific performance. 
While our work shares a similar ideology of exploiting heterogeneity, we differ in our methods for learning optimal configuration and in the domain specific solution for implementing reconfiguration. 
\theo{There's a lot cited above -- but no a lot of details about their difference. I'd hope that the TCP-ICW can be easily included above. for the video -- we need to say someting about Oboe and pensieve. perhaps penseive can be discussed above in the CC section too}
\usama{addressed oboe below. in have addresses pensieve with pytheus}
\theo{How is BO domain specific? I believe, CherryPick explicitly tried to be somewhat domain agnostic. where is Ottertune? }
\usama{Fixed. ottertune is mentioned above.}

Specifically, related approaches model~\cite{alipourfard2017cherrypick, duan2009tuning,van2017automatic} static workloads where performance metrics do not change with time or explore offline~\cite{sigcomm:oboe}, whereas we propose an online approach to tackle network and workload dynamics.
%
Unlike~\cite{tcpwise, niereducing,liu2016efficiently} which tune one parameter, \sysname tunes multiple parameters simultaneously-- ~\cite{tcpwise, niereducing} tunes a subset of those tuned by \sysname where as \cite{liu2016efficiently} tunes orthogonal parameter(i.e., Routing).
\theo{footprint also tackles some server configuration, which ones?} \usama{Their final decisions are routing related. their system uses metrics like CPU, memory usage and latency but tunes them through routing.}
%

Recent works~\cite{nsdi:pytheas,mao2017neural} attacks an othorgonal space -- video bitrate and CDN selection, whereas we focus on network protocols which support the video transfers.  The domains lead to difference in designs. For example, where as Pytheus focus on simple exploitation/exploration with exploration constrainted to cluster-specific data -- we explore a contextual bandit with multiple exploration arms and our exploitation arm is able to generalize across clusters.


\textbf{CrossLayer Optimizations.} 
 We differ from existing cross-layer optimizations~\cite{bridges2007configurable, chen2016dispersing, al2011overclocking} which introduce APIs to enable the different layers to communicate and react accordingly. Instead, we externalize the optimization logic and present an interface across the different layers to enable an external entity to configure the different layers. 



	\section{Conclusion}
\label{s:conclusion}
In this paper, we argue that ``one-size-fits-all'' approach to configuring web serving stacks results in sub-par performance for end-users, especially those in emerging regions.
This argument stands in stark contrast to the traditional setup of today's web serving stacks where a single configuration is used for a divergent set of users.



This paper takes the first step towards realizing heterogenity and fine-grained reconfiguration in a principled and systematic manner: our system, \sysname, introduces a principled framework for learning better configurations, than default, for a connection by systematically exploring the performance of different configurations across a set of similar connections. We demonstrate the benefits of \sysname using both a live deployment and a large scale simulation.

	\newpage

\end{document}